\documentclass[3p]{elsarticle}

\usepackage{verbatim}
\usepackage{graphicx}
\usepackage{array}
\newcolumntype{L}[1]{>{\raggedright\let\newline\\\arraybackslash\hspace{0pt}}m{#1}}
\usepackage{longtable}
\usepackage{lscape} 
\usepackage{hyperref}
\usepackage{threeparttable}

\hypersetup{
	colorlinks=true,
	citecolor= blue,
	linkcolor=red,	
	urlcolor=cyan,
	bookmarks=true}

\usepackage{booktabs}
\usepackage{tabularx}
\usepackage{tikz}
\usepackage{forest}
\usepackage{amsmath}
\usepackage{amssymb}
\usepackage{mathtools}
\usepackage{multirow}
\usepackage{caption}
\usetikzlibrary{arrows.meta}
\usepackage[symbol]{footmisc}
\usepackage{rotating, graphicx}
\usepackage{footnote}
\usepackage[linesnumbered,ruled,vlined]{algorithm2e}

\newcolumntype{L}[1]{>{\raggedright\let\newline\\\arraybackslash\hspace{2pt}}m{#1}}

\author{Unaiza Sajid}
\author[focal]{Rizwan Ahmed Khan}

\author{Shahid Munir Shah}
\author{Sheeraz Arif}
\cortext[cor1]{Corresponding author: Rizwan17@gmail.com}

\address{Department of Computer Science, Faculty of Information Technology, Salim Habib University, Karachi, Pakistan \\

rizwan17@gmail.com; rizwan.khan@shu.edu.pk; shahid.munir@shu.edu.pk; sheeraz.arif@shu.edu.pk

}

\begin{document}

	\begin{frontmatter}
		\title{Breast Cancer Classification using Deep Learned Features Boosted with Handcrafted Features}


\begin{abstract}
	
Breast cancer is one of the leading causes of death among women across the globe. It is difficult to treat if detected at advanced stages, however,  early detection can significantly increase chances of survival and improves lives of millions of women. Given the widespread prevalence of breast cancer, it is of utmost importance for the research community to come up with the framework for early detection, classification and diagnosis. Artificial intelligence research community in coordination with medical practitioners are developing such frameworks to automate the task of detection. With the surge in research activities coupled  with availability of large datasets and enhanced computational powers, it expected that AI framework results will help even more clinicians in making correct predictions.  In this article, a novel framework for classification of breast cancer using mammograms is proposed. The proposed framework combines robust features extracted from novel Convolutional Neural Network (CNN) features with handcrafted features including HOG (Histogram of Oriented Gradients) and LBP (Local Binary Pattern). The obtained results on CBIS-DDSM dataset exceed state of the art.

\end{abstract}

\begin{keyword}
\texttt{Breast Cancer Analysis \sep Mammogram \sep Machine Learning \sep Artificial Intelligence \sep Deep Learning \sep  Medical Imaging \sep Convolutional Neural Network}
\end{keyword}

\end{frontmatter}

\section{Introduction} \label{intro}

Prevalence of different types of cancers is one of the major public health concerns. According to IARC (International Agency for Research on Cancer), a sub committee of the WHO (World Health Organization) breast cancer is the most prevalent cancer (2.1 million new cases in 2018) in women and is the leading cause of cancer deaths in women globally (627 000 deaths in 2018) \cite{cancStat}. 268,600 cases of breast cancer were reported in 2019 in the United States, which is a record figure \cite{desantis2019breast, man2020classification}.

\begin{figure}[htb!]
	\centering
	\includegraphics[scale=0.6]{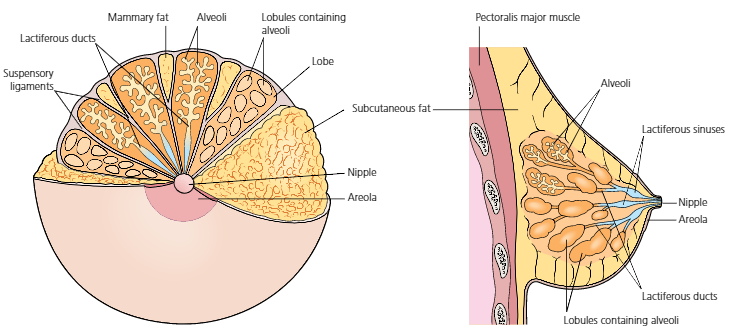}
	\caption {Section of the breast: (left) frontal view showing inner structure of the mammary gland; (right) side view of the breast \cite{bAnatomy}.}
	\label{fig:ana}
\end{figure}

The anatomy of breast comprises of  various connective tissue, blood vessels, lymph nodes, and lymph vessels. The anatomy of the female breast is shown in Figure \ref{fig:ana}. Breast cancer is linked to abnormal and excessive, disorganized, and invasive  division and growth of ducts or lobules cells. This growth causes development of tumor. Tumor is either benign or malignant. Benign tumor or lumps are noncancerous. Malignant tumors are cancerous and they spread through the blood or lymphatic system  \cite{chiao2019detection, 9614200}.

 Malignant tumors are further categorized as invasive (invasive carcinoma) or non invasive (in-situ carcinoma) \cite{chiao2019detection, mahmood2020brief}. Invasive  tumors spread into surrounding organs \cite{cruz2017accurate}.  Invasive Ductal Carcinoma (IDC) (a type of invasive tumor) causes the highest number of deaths in women \cite{BCS2013}.  Non-invasive tumors do not invade neighboring organs and stay in the damaged cell \cite{SHAH2022105221}. Lobular carcinoma in-situ (LCIS) and ductal carcinoma in-situ (DCIS) are the two most common categories of  non-invasive breast tumors \cite{niBrCar}.

Various statistical studies have concluded that breast cancer incidence rate is lower in low income countries / regions as compared to high income countries / regions \cite{BrCaseInci}.  Some studies have suggested that mortality rate in high income countries is lower despite higher incidence rate and vice versa \cite{9614200}. This trend reflects on the socioeconomic factors as one such factor for higher mortality rate in low income countries is lack of access to medical facilities \cite{ASCO}. Other factors that impacts on the mortality rate are age of the subject, ethnicity, reproductive patterns, hormonal and environmental factors, and alcohol and tobacco consumption \cite{Kamiska2015}. It is established that chances of survival depends on the type / sub-type of cancer and stage of cancer at detection. Cancer detected at later stages or when tumor has significantly impacted surrounding organs is difficult to treat.

Early detection and classification is the key to successful treatment of breast cancer. It has been found out that detection and classification of breast cancer at early stages can reduce the mortality rate from 40\% to 15\% \cite{WINTERS20171}. Conventionally clinicians detect breast cancer by employing the triple assessment test, known as ``the gold standard'' \cite{VAKA2020320, BrCML}. Three assessment techniques are: 

\begin{enumerate}
	\item Clinical examination
	\item Radiological imaging
	\item Pathology (Fine Needle Aspirate Cytology (FNAC) or core needle biopsy)
\end{enumerate}

Clinicians analyze all three results to come up with the final conclusion \cite{ BrCML}. All these steps are time consuming and require extreme focus and expertise of medical practitioners. Assessment of radiologists on these radiological images, without the help of any Computer-Aided Diagnosis (CAD) systems, is prone to error (human) and also limits the potential of radiologists as only limited number of images can be examined in a certain time period. 

Other complications that arise due to manual examination of radiological images are \cite{SHAH2022105221}: 
\begin{enumerate}
	\item Shortage of experts with sufficient domain knowledge
	\item Cumbersome activity can lead to increase in false prediction
	\item Remote areas usually lack such expertise
	\item Some tumors in early stages are difficult to get detected and classified 
\end{enumerate}

 Recently, Computer Aided Diagnosis (CAD) systems \cite{doi2005current, giger2008anniversary, sadaf2011performance} have become an essential tool for the pathologists and radiologists for early detection, classification and prognosis of tumors \cite{nishikawa2010computer}. One such example of CAD system for the detection and classification of lung nodules, breast cancer and brain tumor is presented in \cite {jimenez2020deep}. 
 
 In essence, a CAD system for the analysis of radiological images is Artificial Intelligence (AI) (machine learning based in strict sense) based automated image analysis and interpretation algorithm / tool. CAD system limits human dependency, increases correct prediction rate, and reduces the false-positive rate. A computationally efficient and reliable CAD system is now an urgent need for medical diagnostic processes and is being widely used by the radiologists as a second opinion for early detection and classification of tumors \cite{gandomkar2016computer}.

 CAD systems can be divided into two categories 1) Computer-Aided Detection (CADe) and 2) Computer-Aided Diagnosis (CADx). CADe detects and localizes the affected regions in the medical imaging modalities (radiological images e.g. mammograms, ultrasound, magnetic resonance imaging (MRI), histopathology images etc), whereas,  CADx performs the characterization (i.e., normal, benign, and malignant) of lesion present in the human body. Both the systems together make the overall CAD system more important in identifying the abnormality at earlier stages.

As mentioned above, state of the art CAD systems take advantage of the AI algorithms to automate the task of radiological breast image / medical image analysis \cite{talo2019automated, george2020breast, rodriguez2019stand}. Recent wave of successful AI systems leverages on the robust algorithms from the domain of Machine Learning (ML) which is a sub-domain of AI. Broadly, ML algorithms can be classified into two categories: 1) algorithms that use handcrafted features from the given data, 2) algorithms that process raw data to extract features / information that help to classify data points. First category of algorithms are termed as conventional ML algorithms, whereas second category of algorithms are comparatively new and have surpassed results achieved with conventional ML algorithms. Deep Learning (DL)  / Deep Neural Networks (DNN) belongs to the second category \cite{MLDL}.

In this article, we have proposed a novel and robust framework for breast cancer classification using mammogram images, refer Section \ref{NA}. Novel framework uses hybrid features e.g. handcrafted features concatenated with features extracted from novel Convolutional Neural Network (CNN). Refer Section \ref{dlf} for discussion on novel CNN architecture. Extracted handcrafted features are discussed in Section \ref{hcf}. Novel framework has achieved results that exceeds state of the art. Discussion on results is presented in Section \ref{framRes}. The framework is tested on the dataset containing mammograms. The dataset is called  CBIS-DDSM (Curated Breast Imaging Subset of DDSM) \cite{cbis}. Discussion on the dataset is presented in Section \ref{ds}. Classification results obtained using transfer learning approach are used as baseline results. Section \ref{method} presents details related to transfer learning approach and experimental results. Lastly, conclusions are presented in Section \ref{conc}.

\section{State of the art}\label{sota}

Different domains are leveraging the power of AI / ML to maximize efficiency and output by taking data driven decisions \cite{AImark}. Some of the hurdles that are removed in the last fifteen years that lead to its widespread utilization are: advancement of compute power i.e High Performance Computing (HPC) \cite{potok2018study}, open access to large volumes of data, availability of sophisticated algorithms and surge in research activities due to interest of funding agencies \cite{SIVARAJAH2017263, SHAH2022105221, shah2020secondary}.

 AI / ML techniques are not only specific to healthcare domain, but they are also used in almost all the domains, including network intrusion detection \cite{NAZIR2021102164}, image synthesis \cite{Crenn2020}, optical character recognition (OCR) \cite{M2m2020}, facial expression recognition \cite{KHAN20131159} etc. In the healthcare domain various applications are getting benefited from the AI / ML techniques, like remote patient monitoring, virtual assistance, hospital management and drug discovery \cite{Jalia2020, yu2018artificial},  radiological imaging / analysis related tasks i.e.  risk assessment, disease detection, diagnosis or prognosis, and therapy response \cite{giger2018machine} etc.

As mentioned in Section \ref{intro}, there are two categories of ML algorithms : 1) algorithms that use handcrafted features from the given data (conventional ML algorithms) 2) algorithms that process raw data to extract features / information that help to classify data points (deep learning). Much of the early success in the development of CAD system was based on conventional ML algorithms.  In conventional ML algorithms, domain expertise is utilized to extract meaningful information e.g. features from the data, before applying classification technique \cite{KHAN20131159}. Most of these features can be categorized into 1) appearance features, 2) geometric features, 3) texture features, and 4) gradient features \cite{5478947, BERBAR201863, tang2019role, Khan10100}.

In a review by Yassin et al. \cite{yassin2018machine} commonly used conventional ML algorithms / classification techniques employed in the recent past for the breast cancer diagnosis are discussed. These algorithms include Decision Tree (DT) \cite{quinlan2014c4}, Random Forest (RF), Support Vector Machines (SVM) \cite{vapnik1999nature}, Naive Bayes (NB), K-Nearest Neighbor (KNN) \cite{knncc}, Linear Discriminant Analysis (LDA), and Logistic Regression (LR) \cite{agarap2018breast, sharma2017machine}.

Before the advent of deep learning (DL) \cite{lecun2015deep}, most of the research activities have been carried out to propose robust features for breast image analysis, while taking domain expertise into consideration. DL algorithms with appropriate training learns data representation / robust features without human / expert intervention, from the given dataset that are suitable for classification task in hand. DL algorithms merge feature extraction and classification steps in one block, usually a black box \cite{BB}. DL algorithms have achieved state of the art results for breast cancer detection and classification using different breast imaging modalities, e.g. Magnetic Resonance Imaging (MRI), Positron Emission Tomography (PET), Computed Tomography (CT), Ultrasound (US)  and Histopathology (HP) \cite{cheng2016computer,litjens2016deep,todoroki2017detection, murtaza2019deep}.

Most of the DL techniques that have achieved state of the art result on visual stimuli e.g. images and videos,  are based on Convolutional Neural Network (CNN) \cite{lecun2010convolutional, 726791, alexNet}. CNNs are a class of deep, feed forward neural networks that are specialized to extract robust features from visual stimuli \cite{KHAN201961}. As the name suggests, CNNs  uses convolution operations which are basic building blocks for image analysis. Thus, they are very useful in analyzing breast images for the detection and classification of cancer as well \cite{yari2020deep}.

\begin{figure}[!htb]
	\centering
	\includegraphics[scale=0.5]{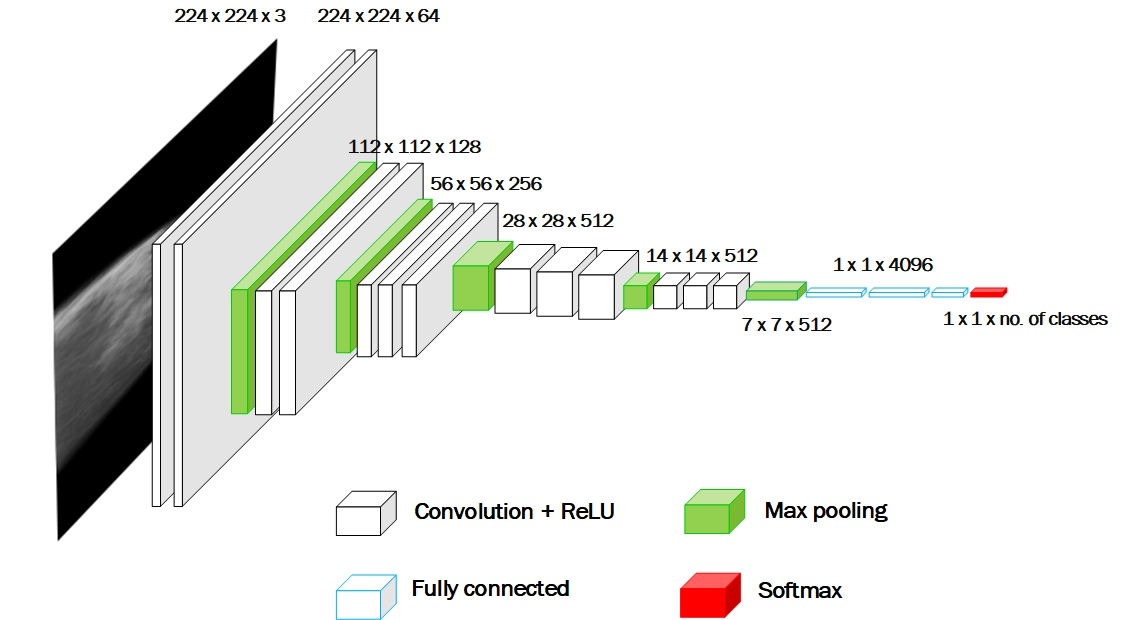}
	\caption { An illustration of VGG16 architecture \cite{simonyan2014very}.}
	\label{fig:vgg}
\end{figure}

The architecture of CNN was initially proposed by LeCun \cite{lecun2010convolutional}. It has multi-stage or multi-layer architecture. Every layer produces an output known as feature map which then becomes an input to the next layer. Feature map consists of learned patterns or features. Initial layers extract low-level features e.g. edges, orientations etc. and layers near to the classification block learns high-level features \cite{szegedy2017inception}. Refer to Figure \ref{fig:vgg} and Figure \ref{fig:Novel Comapct VGG16} for reference.


Next, overview of few seminal works using CNNs and DDSM \cite{ddsm} or CBIS-DDSM \cite{cbis} datasets for breast cancer classification is given. Different articles are cited in chronological order to get better understanding of the direction of research works.  We outlined studies using these datasets as in this study we have also used the same dataset i.e. CBIS-DDSM which is derived from DDSM (refer Section \ref{ds} for discussion on the dataset). Carneiro et al. \cite{carneiro2015unregistered} used pre-trained CNN architecture (see Section \ref{method} for discussion on CNN training methods), resized images of the dataset to $264 \times 264$ and achieved AUC of 0.91.

Suzuki et al. \cite{suzuki2016mass} used Transfer Learning (TL) approach (see Section \ref{method} for discussion on CNN training methods). They adopted Deep CNN (DCNN) AlexNet architecture \cite{krizhevsky2012imagenet} with five convolutional layers and three fully connected layers. In the transfer training phase, authors used 1,656 ROI images including 786 mass images and 870 normal images. In the test phase only 99 images with mass and 99 normal images were used. Proposed framework achieved mass sensitivity detection of 89.9\% and the false positive was 19.2\%.

Jiao et al.  \cite{jiao2016deep} used interesting technique of merging deep features extracted from different layers of trained CNN architecture to classify breast mass. In order to have enriched representation mid level features were taken from activation maps of convolutional layer 5 and high level features were taken from fully connected layer 7. On 600 images of size $227 \times 227$, their framework achieved accuracy of 96\%.

Zhu et al. \cite{zhu2018adversarial} proposed end-to-end network for mammographic mass segmentation, followed by a conditional random field (CRF). As the dataset was small, they used adversarial training to eliminate over-fitting.

Ragab et al. \cite{peerjRagab} proposed classification technique for benign and malignant mass tumors in breast mammography images. They also used TL approach with AlexNet architecture \cite{krizhevsky2012imagenet} but last layer, which is responsible for final classification, was replaced with Support Vector Machine (SVM) classifier \cite{vapnik1999nature}. Their proposed architecture achieved accuracy of s 87.2\% on CBIS-DDSM dataset (image spatial resolution was $227 \times 227$).

Singh et al. \cite{SINGH2020112855} worked on conditional Generative Adversarial Network (cGAN) to segment a breast tumor. In this technique, generative network was trained to recognize the tumor area and to create the binary mask, while adversarial network learned to distinguish between real (ground truth) and synthetic segmentation. Then segmentation shape was classified into four classes i.e. irregular, lobular, oval and round with the help of CNN architecture, with three convolutional layers with kernel sizes $9 \times 9$, $5 \times 5$ and $4 \times 4$, respectively, and two fully connected (FC) layers. Their tumor shape classification achieved accuracy of 80\% on DDSM dataset.

Khamparia et al. \cite{khamparia2021diagnosis} used TL approach for the classification of breast cancer. They tried different pre-trained model for the classification of breast cancer, including AlexNet \cite{krizhevsky2012imagenet}, VGG \cite{simonyan2014very}, ResNet \cite{res} and achieved best results on VGG16. Their framework achieved accuracy of 94.3\% and AUC of 93.6.




Finally, our contributions in this study are threefold: 

\begin{enumerate}
	\item Proposing novel DL / Convolutional Neural Network (CNN) architecture for analysis of mammograms.
	\item Presenting a novel framework for the classification of breast cancer. The proposed framework not only uses features extracted from DNN/CNN but also combines these features with handcrafted features which helps in embedding domain expert knowledge. Thus, proposed framework works on robust features that helps in increasing accuracy of classification.  
	\item Showing that transfer learning approach is good for obtaining baseline results. 
\end{enumerate}

\section{Dataset} \label{ds}

\begin{figure}[!htb]
	\centering
	\includegraphics[scale=0.8]{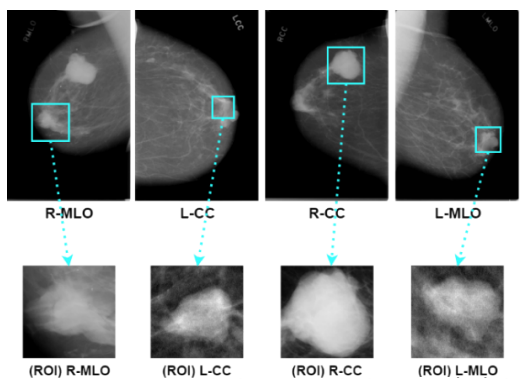}
	\caption {Example Image from the dataset. Mammogram views: CC (the view from above), MLO is an oblique or angled view (taken under 45 degree), Left Cranio-Caudal (LCC), Left Medio-Lateral-Oblique (L-MLO),  Right Cranio-Caudal (R-CC) and Right Medio-Lateral Oblique (RMLO) views of the mammogram \cite{8897609}.}
	\label{fig:cbis}
\end{figure}

For this research, the popular publicly available dataset of breast mammograms CBIS-DDSM is used \cite{cbis}. This CBIS-DDSM (Curated Breast Imaging Subset of DDSM) version of the Digital Database for Screening Mammography (DDSM) is an upgraded and standardized version of the DDSM \cite{ddsm}. The DDSM is a digitized film mammography repository with 2,620 studies. It includes cases that are benign and malignant with their respective pathological information. The DDSM is a useful dataset in the development and testing of CAD systems because of its scalability and ground truth validation.

A subset of the DDSM data was selected and curated by a professional mammographer for the CBIS-DDSM collection. Decompression and conversion to DICOM format were performed on the images, and updated ROI segmentation and bounding boxes are included, with the pathologic diagnosis for training data \cite{cbis}. The data is structured in such a way that each participant has multiple IDs to provide the information of scans of different positions of the same patient. Therefore, it has 6,671 mammogram images in total. All the DICOM images were converted to the PNG format, as well as resized at 255 x 255. The dataset is by default distributed as 80\% training set and 20\% testing set.

\section{Breast cancer classification using transfer learning} \label{method}

In literature, two techniques are applied to train CNN model / architecture, namely Denovo and Transfer Learning (TL).  Denovo as the name suggests, trains complete CNN architecture from the scratch and thus optimally learns features from the given dataset. This technique is used when dataset is large enough to train CNN architecture. 

Second technique to train CNN architecture is TL. Transfer learning allows distributions used in training and testing to be different \cite{5288526}. In other words, TL technique allows to re-train only few layers (usually last layers are re-trained) of pre-trained CNN model in order to adapt it to given problem and dataset. This technique is helpful when the given dataset is not large enough to train robustly all layers of CNN from the very beginning or when the available compute power is not enough  \cite{KHAN201961, liu1999ensemble}.

\subsection{Transfer Learning} \label{tl}


Generally, for TL pre-trained best performing CNN architectures from ImageNet-Large-Scale Visual Recognition Challenge (ILSVRC) \cite{ILSVRC15} are used. ILSVRC  has served as a platform to provide CNN architecture for the task of object recognition.  In past, the best proposed CNN architectures for ILSVRC were AlexNet architecture \cite{krizhevsky2012imagenet}, GoogleNet (a.k.a. Inception V1) from Google \cite{GoogleNet} VGG \cite{simonyan2014very}, ResNet \cite{res} etc. Their pre-trained model are freely available for research purpose, thus making a good choice for transfer learning.

To get the baseline results we used TL approach.  Using limited computational power and relatively small dataset  i.e. CBIS-DDSM \cite{cbis} (refer Section \ref{ds} for a discussion on the dataset), the hyper-parameters are tuned and tested. CNN architectures that have been used / evaluated in this study are explained below.



\subsubsection{VGG16}

The VGG architecture \cite{simonyan2014very} is famous for its simplicity i.e. symmetrical shape. It systematically merges low level features into higher level feature. Its two variants are mostly used in research, architectures with 16 layers (VGG16) and 19 layers (VGG19). In this study, VGG16 architecture is used to classify breast mammograms. It takes an image of the size $224 \times 224$ and $3\times 3$ convolution with a 1-pixel stride and 1 padding. Rectified Linear Unit (ReLU) is used as an activation function, with Softmax classifier at the last layer. Moreover, a Convolution Layer is used, which convolved with the image and produced feature maps. A Feature Pooling Layer is also used to minimize the representation of an image so that the architecture will take less time in training.

\subsubsection{ResNet50}

Deep CNNs have steered successions of innovations regarding image classification tasks. Such networks incorporate all high and low-level features with a non-linear function into a multi-layer manner. The  enrichment of features depends upon the depth of the network. However, stacking more layers led the gradients of the loss function to be almost zero or large error gradients that update a network with the large factor, these problems are referred to as vanishing gradient and exploding gradient respectively, which makes the training challenging and degrade the overall performance of the network. 

Residual networks were introduced to address the above-mentioned problems by using a residual block \cite{res}, with 50 convolution layers. ResNet50 architecture uses residual mappings by using shortcut connections which provide connectivity across layers. That means some layers will have the output of the previous layer with the input of previous convolution blocks as well. These skip connections converge the network more quickly and don’t let the gradient be diminished \cite{res}.

\begin{figure}[h!]
	\centering
	\includegraphics[scale=0.6]{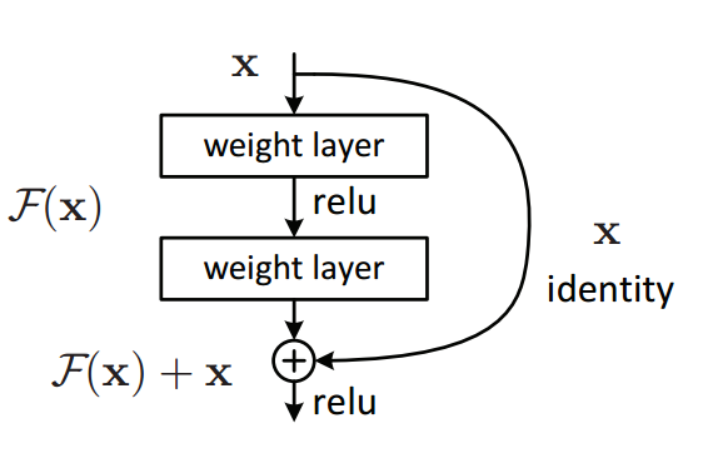}
	\caption{residual block of ResNet50 \cite{res}.}
	\label{fig:resodual block}
\end{figure}

As shown in Figure \ref{fig:resodual block}, the \textit{F(x)+x} is a skip connection, as it is skipping few layers and directly plugging the input into the output of the stacked layer. Skip connections perform identity mapping which neither adds extra parameters nor increases complexity. And hence the whole network can be trained by using stochastic gradient descent.

\subsubsection{DenseNet121}

Recent enhancements have shown that deep convolutional networks are more precise, systematic, and easier to train if there are shorter connections between the layers \cite{dense}. But these connections can lead to the problem of vanishing gradient, similar to the Residual Networks, as they directly store the information by using identity mapping due to which, most of the layers may result in very less or zero contribution. To address this, a dense convolution network was introduced, in which the output of each layer is connected to the input of every other layer in a feed-forward network \cite{dense}.

\begin{figure}[h!]
	\centering
	\includegraphics[scale=0.67]{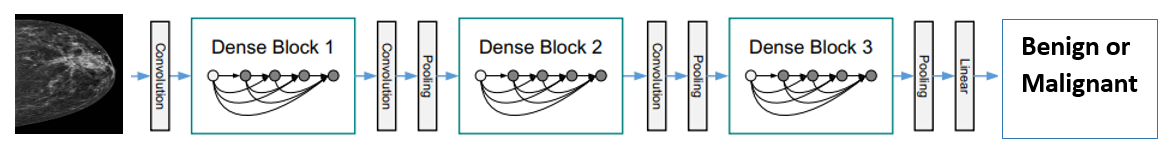}
	\caption{DenseNet with three dense blocks with transition layers in between adjacent dense blocks \cite{dense}.}
	\label{fig:densenet}
\end{figure}

 The feature map of all the former layers is used in all the succeeding layers to obtain collective information \cite{NSR}. As a result, this network topology may be utilized to extract more global and pertinent features, as well as to train with more accuracy and efficiency. Some researchers reprocessed features from all of the layers and then applied these fused features to classify the data. This approach  connects many feature maps, and is not meant to encourage feature reuse between layers. As a result, rather than integrating all feature maps, all prior layers are served as input layers \cite{2019.00080}. Densenet121 consists of multiple blocks of layers that are, the Composition layer, Pooling layer, Bottleneck layer, and Compression layer, as shown in the Figure \ref{fig:densenet}.

\subsubsection{InceptionResnetV1}

In the field of image recognition with the aid of convolution neural networks, inception architectures have performed significantly well with comparatively low cost. The aggregation of residual networks with classic CNN architectures has also shown magnificent results. The researchers in \cite{SzegedyIV}, have developed an ensemble by using the residual blocks with inception architecture to tackle the slow and expensive training of inception architectures. For the combined version of the Inception network with residual blocks, each inception block is followed by a $1 \times 1$ convolution without activation to augment the dimensions. The reason for doing this convolution is to balance the dimensionality reduction introduced by the prior inception block.

\begin{figure}[h!]
	\centering
	\includegraphics[scale=0.65]{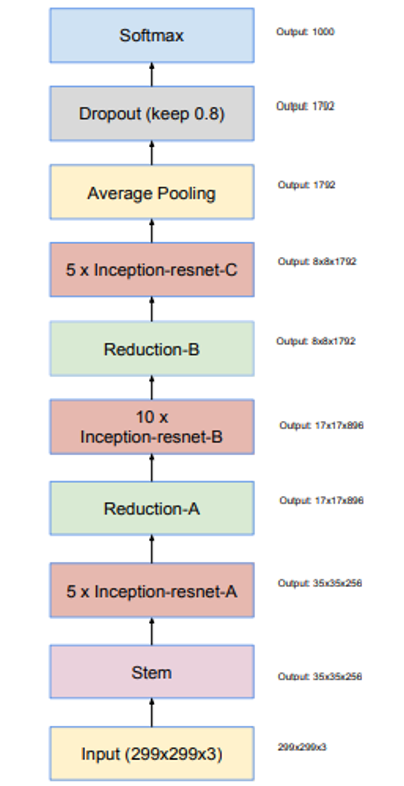}
	\caption{Actual representation for InceptionResnet architecture \cite{SzegedyIV}.}
	\label{fig:inceptionresnet}
\end{figure}

 InceptionResnet incorporates 3 separate stem and reduction blocks. As it is a fusion of inception networks and residual networks, therefore, with the efficiency of inception networks, it has the property of residual networks, that is the output of the inception module is added to the input of the next block.  InceptionResnet schematics overview is shown in Figure \ref{fig:inceptionresnet}. 

\subsubsection{Results and conclusions drawn from transfer learning}\label{resrConcTL}

We applied pre-trained state of the art CNN architectures (that have been discussed in the previous subsections) on CBIS-DDSM dataset \cite{cbis}. The architectures were trained on ImageNet-Large-Scale Visual Recognition Challenge (ILSVRC) \cite{ILSVRC15} dataset, that has more than 1000 classes. We re-trained last layer of these architectures with CBIS-DDSM mammograms.

The  CBIS-DDSM dataset has  6,671 mammogram images in total. All the images were converted to PNG format. The dataset was split with 80\% training set and 20\% testing set.  The results obtained are presented in Table \ref{TabTL}. We also tried to improve these results using different data augmentation techniques e.g. histogram equalization, rotation, flipping etc., but results did not improve. Table \ref{TabTL} presents best results (in terms of accuracy) after optimally tuning hyper-parameters.

\begin{table}[!htb]
	\centering
	\caption{Breast cancer detection and classification by applying transfer learning}
	\label{TabTL}
\begin{tabular}{p{2.7cm} p{1.6cm} p{2.5cm} p{1.5cm} p{1.5cm} p{1.5cm}}
	
	\hline
	Models & Optimizer Function & Loss Function & Learning Rate & Accuracy & Loss \\
	\hline
	
	VGG16 & Adam & Categoical cross-entropy & 0.001 &70\% & 1\%\\
	Resnet50 & Adam & Categoical cross-entropy & 0.01 &69\% & 5\%\\
	DenseNet121 & Adam & Categoical cross-entropy & 0.01 &80\% & 2\%\\
	InceptionResnet & Adam & Categoical cross-entropy & 0.01 & 60\% & 6\%\\

	\hline

	\end{tabular}

\end{table}

		
		
		
		
	

\section{Proposed framework: deep learned features boosted with handcrafted features} \label{NA}

\begin{figure}[!htb]
	\centering
	\includegraphics[scale=0.65]{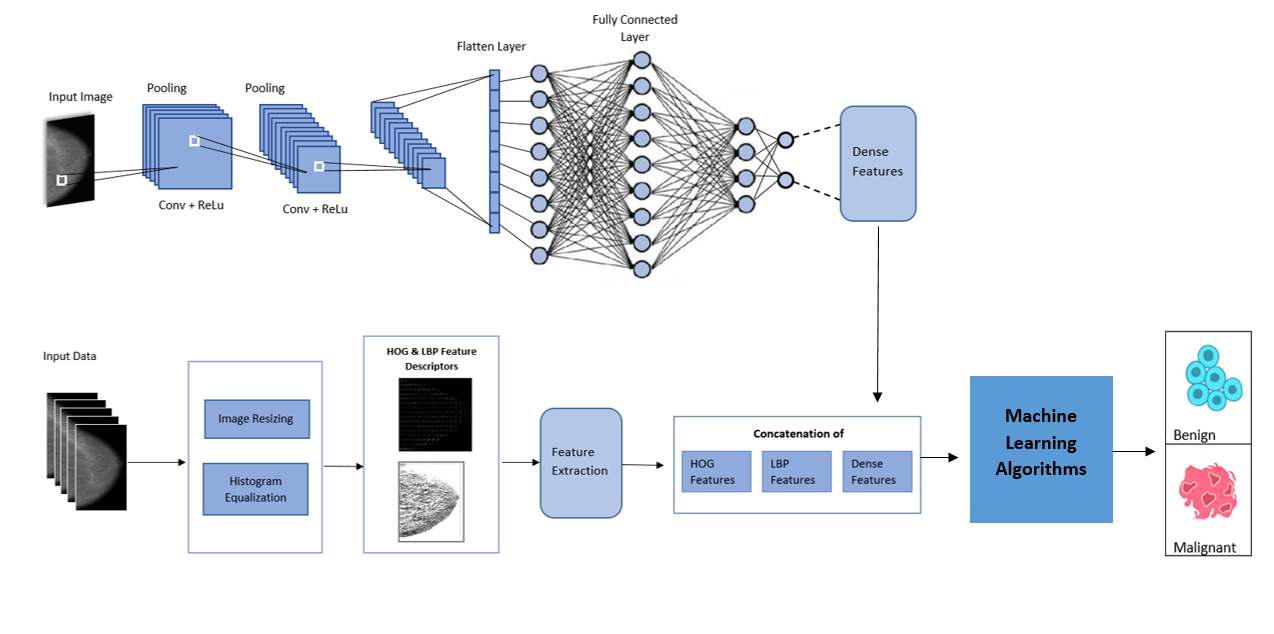}
	\caption{Overview of proposed framework for breast cancer detection and classification that combines deep learned features with handcrafted features}
	\label{fig:Fusion diagram}
\end{figure}

TL approach provides sub-optimal results. This is evident from the results presented in Table \ref{TabTL}, as recognition accuracy is low. Pre-trained architecture employed in TL is trained for much more classes, thus very complex / deep architecture, than number of classes of the problem in hand. In our case, we were dealing with only two classes. With this conclusion, we are proposing a novel architecture for breast cancer detection and classification. This architecture combines deep learned features with handcrafted features. Deep learned features are extracted from the novel CNN architecture, discussed in Section \ref{dlf}. 

Deep learned features extracted using state of the art CNN architecture provides high-level features i.e. recognizable shapes, objects etc. On the other hand, handcrafted features provide low-level information i.e. edge, texture etc, and embeds domain expert knowledge and thus can provide equally important features that can help in the classification task. Discussion on handcrafted features is presented in Section \ref{hcf}. Schematic diagram of proposed framework for breast cancer detection and classification is presented in Figure \ref{fig:Fusion diagram}.

\subsection{Deep learned features} \label{dlf}

\begin{figure}[h!]
	\centering
	\includegraphics[scale=0.65]{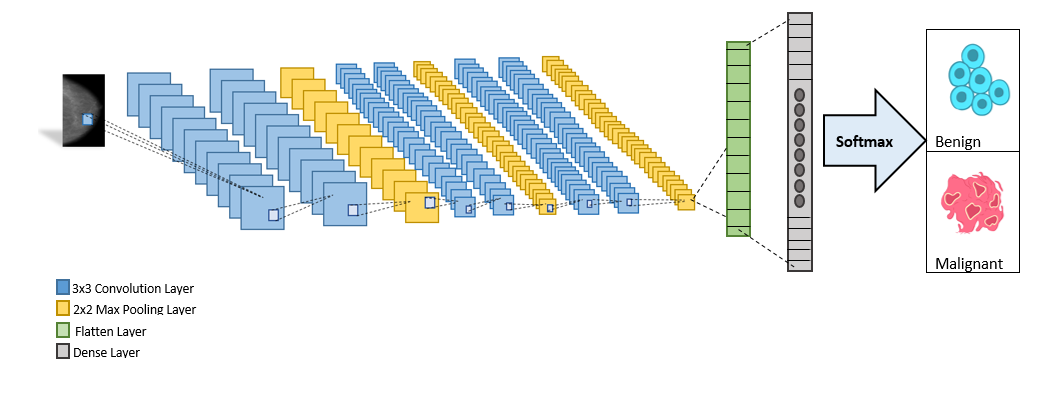}
	\caption{Schematic overview of novel compact VGG (cVGG)}
	\label{fig:Novel Comapct VGG16}
\end{figure}

Since pre-trained VGG16 CNN architecture is trained for the multi-class classification task, which has 1000 different classes. Whereas, breast cancer classification using CBIS-DDSM is a binary classification task, thus utilizing such a deep architecture would be sub-optimal. To cater this drawback, we are proposing simpler, more compact CNN architecture. It is inspired by VGG and thus it is called compact VGG or cVGG. Another reason to be inspired by the VGG architecture is its simplicity (ease of training) and symmetry of data flow, although its accuracy was less than DenseNet121, refer Table \ref{TabTL}.

This variant of VGG16 comprises 12 layers, including a $3 \times 3$ convolution layer, max-pooling layer, a flatten layer, and two dense layers. It has eight layers with trainable parameters. Refer Figure \ref{fig:Novel Comapct VGG16} for schematic overview of cVGG. The activation function used in convolution layers is Rectified Linear Unit (ReLU). This architecture results in approximately 26 million trainable parameters (as opposed to $\approx$ 140 Million trainable parameters for VGG16) with 256 dense features at the last layer. These dense features, containing high-level features, from this novel architecture were then given concatenated with handcrafted features for the final classification.  Details of cVGG are shown in Table \ref{tab:cvgg}. This architecture is empirically found to be optimal for the CBIS-DDSM dataset.



\begin{table}[!htb]
	\centering
	\caption{Structure of novel cVGG}
	\label{tab:cvgg}
\begin{tabular}{ c c c c }
	\hline
	Layer Name& Output Shape & Number of Parameters & Activation Function \\
	\hline
	Conv1 & $253 \times 253 \times 64$ & 1792 & ReLU\\
	Conv1 & $251 \times 251 \times 64$  & 36928 & ReLU \\
	MaxPooling1 & $125 \times 125 \times 64$ & -  & -\\
	
	Conv2 & $123 \times 123 \times 128$  &  73856  & ReLU \\
	Conv2 & $121 \times 121 \times 128$  & 147584 & ReLU \\
	MaxPooling2 & $60 \times 60 \times 128$  & -  & - \\
	
	Conv3 & $58 \times 58 \times 128$ & 147584 & ReLU \\
	Conv3 & $56 \times 56 \times 128$ & 147584   & ReLU \\
	MaxPooling3 & $28 \times 28 \times 128$ & -  & - \\
	
	Flatten & 100352 & -  & - \\
	Dense4 & 256 & 25690368  & ReLU \\
	Dense4 & 2 & 514 & softmax \\
	
	\hline
\end{tabular}
\end{table}

\subsection{Handcrafted features} \label{hcf}

Handcrafted feature extraction techniques process the given stimuli i.e. mammograms, and extract relevant information.  The optimal features minimize within-class variation, while maximize between class (malignant and benign) variations. If sub-optimal features are extracted then ML classifiers fail to achieve robust recognition \cite{KHAN20131159, MUNIR2019102660}. Handcrafted feature allows to embed expert knowledge in the process of feature extraction. 

Wolfe`s parenchymal patterns \cite{w76} showed that texture characteristics of mammograms are important for cancer detection. Wang et al. \cite{wang2017a} study indicated that shape features are also important to analyze mammograms. Considering knowledge from the domain experts, we extracted handcrafted features that cater shape and texture information from mammograms. Shape features are extracted using HOG (Histogram of Oriented Gradients) \cite{1467360} and texture features are extracted using LBP (Local Binary Pattern) \cite{OJALA199651}.  Details of these features are presented in the next subsections.


\subsubsection{Histogram of Oriented Gradients (HOG)} \label{hog}

The HOG feature descriptor emphasizes the structure of the mass in an image / mammogram. It computes the gradient and orientation of edge pixels. These orientations give information related to global shape of an object present in the image. HOG descriptor divides images into smaller sections and for each section, the gradient with its orientation is computed. Lastly, a histogram is created using the gradients and orientations for each region separately, given the name Histogram of Oriented Gradients.

\begin{figure}[h!]
	\centering
	\includegraphics[scale=0.65]{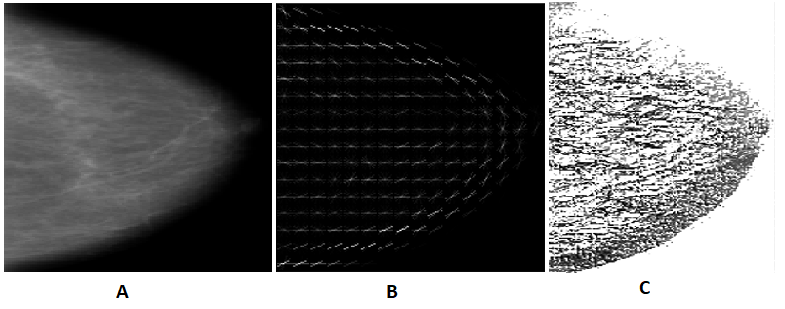}
	\caption{Handcrafted feature extraction from the mammogram. A: Example mammogram from the dataset; B: corresponding HOG features and C: corresponding LBP features. }
	\label{fig:HCFeat}
\end{figure}

 In this study, image of size $255 \times 255$ is passed through the HOG feature descriptor. The descriptor divided the image into several blocks. Each block consists of $16 \times 16$ cells and each cell has a $16 \times 16$ number of pixels that give the resultant vector of 2304 dimensions. Refer Figure \ref{fig:HCFeat} for example image.

\subsubsection{Local Binary Pattern (LBP)}\label{lbp}

The LBP (Local Binary Pattern) feature descriptor is an efficacious method to describe the local spatial patterns and gray-scale contrast in an image. Other important properties of LBP features include tolerance against illumination changes and their computational simplicity \cite{KHAN20131159}. Initial proposed LBP descriptor (which we have also used in this study) works in a $3 \times 3$ window which uses the central pixel as the threshold of the neighboring pixels \cite{OJALA199651} . In a $3 \times 3$ window, for every pixel (x , y) with N neighboring pixels at $ \mathbb{R}$  radius, it computes the difference in the intensities of the central pixel (x,y) with the neighboring N pixels, refer Equation \ref{lbp1}. If the resultant difference is negative, it assigns 0 to the pixel (x,y) or 1 otherwise, creating a bit vector, refer Equation \ref{lbp2}. Lastly, it converts the central pixel value with its corresponding decimal value. Refer Figure \ref{fig:HCFeat} for example image.

\begin{equation}
	LBP (N,R) = \sum_{n=0}^{N-1} f(i_n - i_c)2^p
	\label{lbp1}
\end{equation}

where \({i_n}\) is the intensity of neighboring pixel and \({i_c}\) is the intensity of current pixel. 

\begin{equation}
	f(x) = \Bigg\{1 \:  if x > 0; \: 0 \: otherwise
	\label{lbp2}
\end{equation}




\subsection{Experiment with deep learned features boosted with handcrafted features} \label{resF}

Experiment is performed to find breast cancer detection accuracy on CBIS-DDSM dataset using deep learned features boosted with handcrafted features. The dense features gained from the novel cVGG were concatenated with HOG and LBP features to form one robust feature vector for the task in hand, refer Figure \ref{fig:Fusion diagram}. This concatenation resulted in 2816 dimensional feature vector. Then the performance is evaluated using three different conventional ML classifiers, belonging to different family of algorithms: 

\begin{enumerate}
	\item Random Forest (RF)
	\item $K-$Nearest Neighbor (KNN)
	\item Extreme Gradient Boosting (XGBoost)
\end{enumerate}   

Above mentioned classifiers are briefly described below. 

 


\paragraph{Random Forest}
Random Forest (RF)  falls under the broad category of ensemble learning techniques which involves the development of multiple associated decision trees, used for classification and regression tasks. A classification tree consists of branches and nodes, where each node splits the subset of features according to the associated decision rule \cite{quinlan2014c4}. Random forest is the most widely used Bagging model nowadays for datasets with high variance and low bias. 



\paragraph{$K-$Nearest Neighbor (KNN)}
K Nearest Neighbor belongs to the family of non-parametric, instance-based models that classify  each instance \textit{x} closest to other instances based on the targeted function \cite{knncc}. Given the calculated value of distance metric, it infers the class label of the instance \textit{x}. The probability of an instance \textit{x}, belonging to class \textbf{a} is given by:

\begin{equation} 
	p(a\mid x) = \frac{\sum_{k \epsilon K} W_k .1_{(k_{a}=a)}}{\sum_{k \epsilon K} W_k} 
\end{equation}

where \textbf{K} is the sample space, $ k_a $ is the class label of \textbf{K} samples and \textit{d(k,a)} is the Euclidean distance from \textit{k} to \textit{a}.

\paragraph{Extreme Gradient Boosting (XGBoost)}

XGBoost (Extreme Gradient Boosting) is a robust algorithm that comes under the category of boosting technique \cite{TianqiChen2016}. It is a tree-based ensemble model in which new learners are added to minimize the errors made by the prior learners. XGBoost works similarly to a gradient boosting algorithm that predicts the residuals of prior learners and adds them to the final predicted result. It minimizes the loss of each learner by using a gradient descent algorithm \cite{aoms}. XGBoost is known for its scalability and a sheer blend of both software and hardware optimization that generate successful results in the shortest amount of time with the least utilization of computational power. The objective function of XGBoost is defined as follows:

\begin{equation}
	L(y_i , \hat{y_i}) + \sum_{k} \Omega{f_k}
\end{equation}

where,

\begin{equation}
	L(y_i , \hat{y_i}) 
\end{equation}

is a differentiable loss function that evaluates the difference between the predicted and the targeted label for a given instance. \(\Omega{f_k}\) describe the complexity of the tree \({f_k}\). 

In \cite{aoms}, the objective function of XGBoost is elaborated as:

\begin{equation}
	\Omega({f_k}) = \gamma T + \frac{1}{2}\lambda \omega^2
\end{equation}

where, \textbf{T} is the number of leaf nodes of the tree \({f_k}\) and \(\omega\) is called the predicted values on the leaf nodes (weights). \(\omega{f_k}\) plays an important role in the optimization of a less complex tree and instantaneously  reduces the loss \(L(y_i , \hat{y_i}) \) of the tree \({f_k}\). Over-fitting is reduced by adding a penalty constant for each addition leaf node, \(\gamma\) T and \(\omega^2\) penalizes extreme weights. Here both \(\gamma\) and \(\lambda\) are the hyper-parameters for an optimal tree.

\subsection{Results} \label{framRes}

This study is carried out using the breast mammograms scans from the CBIS-DDSMS dataset, which is one of the widely used datasets for digital mammograms. It has two classes, benign and malignant with the information of 2,620 specimens. It also has four CSV files for the detailed information of each image, which includes, the patient's name, its CC or MLO position, calcification and mass sample, etc \cite{cbis}.

\begin{table}[!htb]
	\centering
	\caption{Accuracy on deep learned features boosted with handcrafted features }
	\label{tab:resFF}
	\begin{tabular}{ l c  }
		\hline
		Classifier & Accuracy  \\
		\hline
		
		Random Forest & 75\% \\
		KNN & 85\%\\
		XGBoost & 91.5\%\\
		\hline
		
	\end{tabular}
	
\end{table}

As discussed above, in this proposed framework we have tried to leverage on the deep learned features from proposed novel CNN architecture and handcrafted features. Both features provide complimentary information as it is evident from Tables \ref{tab:resFF} and Table \ref{tab:resSum} that accuracy substantially improves with the fusion / concatenation of these features. The ensemble classifier, XGBoost achieved accuracy of 91.5\% which exceeds state of the art. Even, naive non-parametric KNN classifier achieves improved recognition accuracy. This could be due to the fact that feature space is well portioned and extracted features have optimal inter-class variance.

To provide relative strength and weaknesses of different modalities, summary of results obtained from different experiments are presented in Table  \ref{tab:resSum}. In Table  \ref{tab:resSum} we have also presented results of deep learned features used in isolation (shown with *section). Results were not robust enough. Results obtained with the fusion / concatenation of features improve accuracy substantially.


\begin{table}[!htb]
	\centering
	\caption{Summary of experimental results}
	\label{tab:resSum}
\begin{tabular}{ l c }
	\hline
	\textbf{Technique} & \textbf{Accuracy} \\ \\
	\hline
	\textbf{Transfer learning} & \\
	\hline
	VGG16 & 70\% \\
	ResNet50 & 69\%  \\
	DenseNet121 & 80\% \\
	InceptionResnet & 60\%\\
	
	\hline
	\textbf{Conventional ML applied on deep learned features*}& \\
	\hline
	Random Forest & 57\% \\
	KNN & 55\% \\
	XGBoost & 60\% \\
	
	\hline
	\textbf{Deep learned features boosted with handcrafted features}& \\
	\hline
	Random Forest & 75\% \\
	KNN & 85\% \\
	XGBoost & 91.5\% \\
	
	\hline
\end{tabular}
\end{table}


\subsubsection{Comparison with the state of the art methods}
 
 We have compared results of the proposed framework with representative methods in the literature that have used DDSM and / or CBIS-DDSM datasets. The result comparison summary is presented in Table \ref{tab:resComp}.

\begin{table}[!htb]
	\centering
	\caption{Comparison with the state of the art that have used DDSM and/or CBIS-DDSM datasets}
	\label{tab:resComp}
\begin{tabular}{p{2cm} p{1cm} p{5.8cm} p{2cm} p{4.5cm}}
	\hline
	
	\textbf{Reference} & \textbf{Year} &\textbf{Features / methodology} &\textbf{Dataset} & \textbf{Accuracy} \\
	\hline
	\\
	
		Ragab et al. \cite{peerjRagab} & 2019 & Classification: mass and normal lesions using TL & DDSM & 87.2\% \\

	Singh et al. \cite{SINGH2020112855} & 2020 & Segmentation,  tumor shape Classification: irregular, lobular, oval and round & DDSM  & Shape Classification accuracy: 80\% \\
	

	Salama et al. \cite{SALAMA20214701} & 2021 & Segmentation and classification using state of the art CNN architectures (TL approach) & MIAS, DDSM, and CBIS-DDSM & TL on:
	\begin{enumerate}
		\item VGG16: 80.98\%
		\item DenseNet121: 82.47\%
		\item MobileNetV2: 79.82\%
		\item ResNet50: 81.65\%
		\item InceptionV3: 84.21\%
	\end{enumerate}

	\\

	
	Tsochatzidis et al. \cite{TSOCHATZIDIS2021105913} & 2021 & Segmentation and Diagnosis Mammographic Mass. Modified pre-trained ResNet50 and modified loss function to focus on mass region & DDSM (400) and CBIS-DDSM (1837) &  AUC 0.89 (DDSM), 0.86 (CBIS-DDSM) \\

	Khan et al. \cite{10.1371_khan} & 2021 & Classification: Calcification, Masses, Asymmetry and Carcinomas. Combination of ResNet50 based TL approach combined with novel CNN &               CBIS-DDSM & 88\% \\

	
	\\
	\textbf{Ours} & \textbf{2022}  & \textbf{Deep (learned from novel CNN architecture) and hand-crafted (HOG + LBP) features fusion} & \textbf{CBIS-DDSM} & \textbf{91.5\%} \\
	\hline
	
\end{tabular}
\end{table}

Table \ref{tab:resComp} provides a quick overview of how the field is progressing in terms of methods utilized and accuracy. Our proposed method is simple yet robust enough to achieve accuracy that exceeds state of the art.

\section{Conclusion and future directions} \label{conc}

There is no denying the fact that field of AI has impacted many fields that also includes the domain of medicine. Specifically, there is a surge in AI based systems that analyzes medical images to detect many diseases, including breast cancer. 

In this article we have presented a novel framework aiming to detect breast cancer by analyzing mammograms. Initially, we gathered baseline results using transfer learning applied on state of the art CNN architectures. To  remove drawback of transfer learning approach, we proposed novel CNN architecture  i.e. cVGG. This novel CNN architecture was trained on mammograms with the aim to classify them either as malignant or benign. To boost results obtained from the novel CNN architecture, handcrafted features were introduced keeping in view that these features will provide complimentary information that embed knowledge from domain experts.

Following conclusions were drawn from this study: 
\begin{enumerate}
	\item Transfer learning approach is good to obtain baseline results. 
	\item Result obtained from transfer learning can be improved.
	\item Complexity of CNN architecture doesn't guarantee robust results. 
	\item Fusion of low-level features with high-level features can improve recognition accuracy. 
\end{enumerate}

Although researchers working in the domain of AI have proposed many frameworks / solutions for the detection and classification of breast cancer but clinicians and physicians are not actively adopting these solution due to the very nature of framework i.e. ``black-box''. Deep learning / CNN architectures are  black-box in nature i.e. it is difficult to know the reasoning or justification of prediction made by the AI algorithm. However, for clinicians, it is important to understand the reason that led to any particular decision \cite{jam2010}. Thus, explainability of the model is important aspect for the technology adoption \cite{imSci2017}.

Another issue that hinders the development of AI based systems is the availability of comprehensive and fully labeled / annotated
datasets that comply with ethical and international distribution guidelines. Deep learning models are data hungry and need volumes of data to be trained robustly.

\end{document}